\documentstyle[12pt]{article}
\newcommand{\x}{\mbox{$X^{\mu}$}}
\newcommand{\y}{\mbox{$Y^{\mu}$}}
\newcommand{\ps}{\mbox{$\psi_{r}$}}
\newcommand{\nd}{\partial }
\newcommand{\db}{\overline{\nd }}
\newcommand{\nn}{\nd^{2} X}
\newcommand{\bb}{\db^{2} X}
\newcommand{\nb}{\nd \db X}
\newcommand{\keindex}[1]{\partial #1^{\mu}
                         \overline{\partial} #1_{\mu}}
\newcommand{\ke}[1]{\partial #1\cdot \overline{\partial} #1}
\newcommand{\kes}[1]{\frac{#1}{(\ke{X})^{2}} }
\newcommand{\seki}[1]{\int d^{2} #1}
\newcommand{\pse}{\frac{\partial ^{2} f}{\partial \psi_{r} \partial
\psi_{r}}}
\newcommand{\pf}[1]{\frac{\partial #1}{\partial \psi_{r}}}
\newcommand{\ap}{\alpha '}
\newcommand{\pa}[1]{\frac{1}{#1 \pi \ap }}

\def\fnote#1#2{\begingroup\def\thefootnote{#1}\footnote{#2}
    \addtocounter{footnote}{-1}\endgroup}
\def\sppt{Research supported in part by the Robert A. Welch
Foundation, NSF Grant PHY 9009850, and
the Texas Advanced Research Program}

\begin{document}

\begin{flushright}
  UTTG-10-92 \\
  \today
\end{flushright}
\vspace{24pt}
\begin{center}
{\bf Nonlinear Sigma Model for String Solitons}

\vspace{36pt}
Makoto Natsuume\fnote{*}{\sppt}

\vspace{6pt}
Theory Group \\ Department of Physics\\
University of Texas \\ Austin, Texas 78712 \\
bitnet: makoto@utaphy
\vspace{48pt}

\underline{ABSTRACT}

\end{center}

\vspace{24pt}
It is shown that a renormalizable nonlinear sigma model gives
rise to the effective string theory proposed by Polchinski and
Strominger.
In the presence of long string background, the
model contains massive world-sheet degrees of freedom owing to
the spontaneous breaking of
conformal invariance.
\vfill

\newpage
One of the unsolved problems in QCD is the string hypothesis;
that QCD may be reformulated as a string theory. Recently,
the effective string
theory which governs the behavior of a long QCD flux tube has
been proposed by Polchinski and Strominger
(P-S hereafter) \cite{ps}. The basic puzzle they struggled
with is the paradox that none of the
standard string quantizations works for the flux tubes. The
paradox is due to the discrepancy of the number of massless
world-sheet degrees of freedom required by the quantizations
from the $D\! -\! 2$ degrees of freedom required by the symmetry of
the flux tube ($D$ is the dimension of spacetime although our
interest is in four dimension.).
In those quantizations, $D\! -\!2$ is not sufficient number
of degrees of freedom unlike P-S theory which contains
only $D\! -\! 2$
massless world-sheet oscillators at low energy.
For instance, standard bosonic
string theory is covariant only in 26
dimensions, not in 4 dimensions,
in light cone quantization \cite{gg}.

Their theory is, however, an effective theory, so
nonrenormalizable and not valid at sufficiently high energies. It
is therefore
plausible that the flux tube acquires additional massive degrees
of freedom as one increases
the energy. The appearance of additional fields may be considered
as offset of less number of
world-sheet oscillators at low energy than in any standard
quantizations.

The necessity of such massive degrees of freedom has been also
suggested by the work of Germ\'{a}n, Kleinert, and Lynker
\cite{gkl}. They calculated the deconfinement temperature of SU(3)
lattice gauge theory and various string models. They found that
although pure Nambu-Goto string theory gives too high of a value
in comparison with the SU(3) value, the presence of massive scalar
fields in NG string theory shifts the temperature, which brings
it to the SU(3) value.

In this paper, we propose a renormalizable string theory
which is consistent with P-S theory. The theory is
a nonlinear sigma model in $D\! +\! N$ dimensions.
Among the $D\! +\! N$ world-sheet
fields $X^{i}\ (1 \leq i\leq D\! +\! N)$, $D$ of them represent
spacetime coordinates \x\ and the rest
$\psi^{r} (1 \leq r \leq N)$ are the internal degrees
of freedom of this model: $X^{i}=\{ \x,\psi^{r} \}$. The spacetime
and internal degrees of freedom are distinguished by
symmetry: we impose only $D$ dimensional Poincar\'{e} invariance.
We then find that, rather generically, when we expand around
a string which is long in the spacetime direction, the internal
fields are massive. So, at high energy, we have an ordinary $c=26$
nonlinear sigma model. But at low energy, we have a Poincar\'{e}
invariant string theory with only $D$ \x\ fields, where
$D-2$ are dynamical.
This is the same paradox examined by P-S,
and by explicitly integrating out the massive fields we find
that it is resolved as they proposed: their effective Lagrangian
appears at low energy.

This is interesting as a ``derivation'' of the P-S theory in one
example, and also as an example of a conformal field theory with
massive world-sheet fields. World-sheet masses are possible in a
non-critical string, where conformal invariance is spontaneously
broken, but they have been little explored. On the other hand,
we also
find a pathology: an inevitable world-sheet tachyon. Polchinski
has argued that such a pathology is likely to be a rather general
problem
\cite{p1}.

The motivation to use a nonlinear sigma model is to introduce
self-interaction terms for $X^{i}$. Conformal invariance is often
considered as the most important symmetry on string
actions.
It is by no means a principle, but it is commonly believed that
the string theory without this symmetry seems hard to
make sense of since it is just the residual
symmetry from coordinate invariance. But then, massive fields are
not allowed unless a symmetry breaking mechanism through
the interactions among world-sheet fields, which are not present
in the original bosonic string theory, is introduced.
In the nonlinear sigma model, the symmetry breaking is possible
since the operator $\keindex{X}$ has a non-zero
classical expectation value for long strings. In order to
compare our model with P-S theory, let us use the same
classical solution as theirs \cite{ps}:
\begin{equation}
\x_{cl}=R(e^{\mu}_{+}z+e^{\mu}_{-}\overline{z}),
\end{equation}
where $e_{+} \cdot e_{-}=-1/2$. $2 \pi R$ is the string
length since they assume the toroidal compactification of
the closed strings on a circle of radius $R$ with the winding
number 1 to avoid infrared
divergence. The solution is obtained
from the ordinary bosonic string
action, which will be our leading order action.
Note $< \keindex{X}>=-R^{2}/2$.

Now, the nonlinear sigma model action is given by \cite{gsw}
\begin{eqnarray}
\lefteqn{S_{nls}=\seki{\sigma} \pa{4} \{ \sqrt{g} g^{ab}
           G_{ij}(X^{i}) +\epsilon^{ab} B_{ij}(X^{i}) \}
           \nd_{a} X^{i}\nd_{b} X^{j} }\hspace{1in}\nonumber \\
           & &\hspace{1.5in}+\frac{1}{4\pi} \sqrt{g} R^{(2)}
           \Phi (X^{i}),
\end{eqnarray}
where $G_{ij}(X^{i})$ and $B_{ij}(X^{i})$ are respectively
symmetric and antisymmetric tensors and
$R^{(2)}$ is the world-sheet
Ricci scalar.  As mentioned earlier, we impose Poincar\'{e}
invariance for $D$ components of
$X^{i}$. Standard power counting for renormalizability
then restricts the possible form of the tensors:
\begin{eqnarray}
\lefteqn{S=\seki{\sigma} \pa{4} \sqrt{g} g^{ab}
             \{ f(\psi)\nd_{a} X^{\mu} \nd_{b} X_{\mu}
             +g_{rs}(\psi)\nd_{a} \psi^{r} \nd_{b} \psi^{s} \}
}\hspace{1in}\nonumber\\
             & &\hspace{1.5in}+\frac{1}{4\pi} \sqrt{g} R^{(2)}
             \Phi (\psi^{r}) .
\end{eqnarray}
$f$ and $g_{rs}$ are at this point arbitrary and $g_{rs}$
is a symmetric
tensor. We expand $f$ in a Taylor series around
the stationary point. By suitable field redefinitions
of \x\ and \ps\ ,
\[
\begin{array}{ll}
  f(\psi)=1+\frac{1}{R^{2}} \sum_{r=1}^{N}m_{r}^{2} (\psi_{r})^2
         +O(\psi^3) &
  \mbox{with $ \frac{m_{r}^{2}}{R^{2}} \equiv
         \frac{1}{2} \partial_{r} \partial_{r} f |_{\psi =0}$ }.
\end{array}
\]
Here, we also let
$g_{rs}(\psi)=\delta_{rs}$ since the main motivation for
the nonlinear sigma model is to introduce mass terms for
\ps . Since $< \ke{X} > \neq 0$, \ps\ shall
acquire a mass $m_{r}$ by the first term in (3).
To make the treatment simple, we henceforth assume a linear
dilaton background: $\Phi (\ps )=a_{r} \ps $.

After the standard gauge fixing,
\begin{equation}
S=\pa{2} \seki{z} \{ f(\psi)\ke{X}+\ke{\psi} \} +S_{ghost}.
\end{equation}
If \x\ and \ps\ are approximately free ($f \simeq 1$), \x\
and \ps\ contribute central charges as follows \cite{pb}:
\[
\left \{ \begin{array}{l}
        c_{X}=D \\
        c_{\psi} =N+6\ap a_{r}a_{r}.
       \end{array} \right.
\]
The requirement that conformal
invariance not be anomalous, $c_{X}+c_{\psi}=26$ gives $6\ap
a_{r}a_{r}=26-(D\! +\! N)$.

The functional form of interaction, $(f-1) \ke{X}$ is also
constrained by conformal invariance to
be a (1,1) tensor. A standard calculation of OPE gives
\begin{equation}
-\frac{1}{2} \pse +a_{r}\pf{f} =0,
\end{equation}
where the free propagator $- \ap \ln (z\overline{z} )/2$
has been used. Since the free
propagator is used in the calculation, one may expect
it corresponds to the expansion of the metric around a flat one
in target space. In fact,
equation (5) can be obtained from the vanishing condition
of the beta functional in
the nonlinear sigma model (which is implied by
Weyl invariance) by demanding that $f$ be small in the sense that
$| \partial_{r} f | \ll 1$. The appropriate beta functional is
\cite{gsw}
\begin{equation}
\ap R_{ij}-2 \ap \nabla_{i} \nabla_{j} \Phi + O(\ap^{2})=0.
\end{equation}
where $R_{ij}$ is the `spacetime' Ricci tensor. For our choice of
the spacetime metric $G_{\mu \nu}$, the equation (6) gives
\begin{equation}
-\frac{D-2}{4f} \pf{f} \pf{f} -\frac{1}{2} \pse +\pf{f}
\pf{\Phi} \simeq 0 .
\end{equation}
Note the first term in (7) is absent in (5).

Equation (5) reduces to recursion relations for the
coefficients of $f$. The lowest one gives
$\sum_{r=1}^{N} m_{r}^{2}=0$. Thus, $N \geq 2$ to make \ps\
massive. Also, the theory can not be stable due to world-sheet
tachyons. Even though the existence of tachyons is a drawback,
one can continue the analysis formally by analytically continuing
amplitudes as in bosonic string theory.

Once $m_{r}^{2}$ is determined, the recursion relations
constrain higher expansion coefficients in $f$. There is one
particularly simple solution: $N=26-D$ and $f=$ quadratic only.
For the purpose of explicit calculation, we take this example
henceforth.

We now expand our action around the long string vacuum solution
in terms of the fluctuation field $\y =\x -\x _{cl}$.
Except for total derivative terms,
\begin{equation}
S=\pa{2} \seki {z} \{ \ke{Y} +\ke{\psi}
  -\sum_{r=1}^{N} m^{2}_{r} \psi_{r}^{2} +{\cal L}_{int} \} ,
\end{equation}
where
\begin{equation}
{\cal L}_{int}=\frac{2}{R^2} \sum_{r=1}^{N} m^{2}_{r}\psi_{r}^{2}
\{ \ke{Y} +R(e_{+}\cdot \db{Y} +e_{-}\cdot \nd{Y} )\}.
\end{equation}

The fields \ps\ can be integrated out leaving an effective theory
containing only \y\ fields. Lowest order perturbation
expansions as shown in figure 1 give
the following effective Lagrangian in terms of \x\ :
\begin{eqnarray}
\lefteqn{{\cal L}_{eff}=\pa{2} \ke{X} }\nonumber \\
  & &+\frac{\beta}{4\pi}
  \{ \kes{\bb \cdot \nd{X} \nn \cdot \db{X} }
  + \kes{\nb \cdot \db{X} \nn \cdot \db{X} }\nonumber\\
  & &\hspace{.27in}+\kes{\bb \cdot \nd{X} \nb \cdot \nd{X} }
  + \kes{\nb \cdot \nd{X} \nb \cdot \db{X} } \}
    +O(R^{-3}),
\end{eqnarray}
where $\beta =\frac {26-D}{12}$. This is our main result.
Higher order
loops (figure 2) are suppressed by powers of $R^{-1}$.
The first and second terms in ${\cal L}_{eff}$ gives
the P-S theory with the right coefficients. The
extra terms are proportional to the lowest order field equation
and vanish (up to higher derivative corrections) on shell. These
extra terms do not affect the physics but do, of course,
complicate the calculation, and make it necessary to calculate
the 4-point as well as 2-point amplitudes to evaluate
${\cal L}_{eff}$.

Even though the required low energy behavior can be reproduced
from this model, the model may suffer from the serious
problem discussed by Polchinski \cite{p2}. He considered the
temperature dependence on the spectrum of string
theory and of SU($n$) gauge theory. Based on the analysis,
he suggests that the string theory equivalent to
large-$n$ gauge theory has an effective number of world-sheet
fields which diverges at high temperatures. It therefore
suggests the model we have obtained
may not have enough degrees of freedom at higher energies.

Since it is not practically possible to introduce an infinite
tower of fields on the world-sheet, one had better modify
string theory somehow without relying on further addition of
extra fields. Then, a string theory with finite numbers
of additional fields, like given in this paper, may give
correct low and high energy behavior.
Dirichlet string theory suggested by Green \cite{g} may be
a good starting point.
\vspace{.1in}
\begin{center}
    {\large {\bf Acknowledgements} }\\
\end{center}
\vspace{.1in}

I am grateful to J. Polchinski for having suggested the problem
and for his continuous assistance throughout the work. I also
thank J. LaChapelle for critical reading of the manuscript. \\

\pagebreak

\end{document}